\documentstyle[epsfig,aps,twocolumn]{revtex}

\newcommand{\pol}{\hat{\bf e}}
\newcommand{\rv}{{\bf r}}
\newcommand{\Ev}{{\bf E}}

\newcommand{\Hc}{{\cal H}}

\newcommand{\qv}{{\bf q}}
\newcommand{\kv}{{\bf k}}

\newcommand{\beq}{\begin{equation}}
\newcommand{\eeq}{\end{equation}}
\newcommand{\bea}{\begin{eqnarray}}
\newcommand{\eea}{\end{eqnarray}}

\newcommand{\<}{\langle}
\renewcommand{\>}{\rangle}

\renewcommand{\[}{\left[}
\renewcommand{\]}{\right]}

\begin{document}

\draft
\title{Pulsed Raman output coupler for an atom laser}
\author{J. Ruostekoski$^1$, T. Gasenzer$^2$, and D. A. W. Hutchinson$^3$}
\address{$^1$Department of Physical Sciences, University of Hertfordshire,
Hatfield, Herts, AL10 9AB, UK\\ $^2$Clarendon Laboratory,
Department of Physics, University of Oxford, Oxford OX1 3PU,
United Kingdom\\$^3$Department of Physics, University of Otago,
P.O. Box 56, Dunedin, New Zealand}
\date{\today}
\maketitle
\begin{abstract}
We theoretically study a pulsed stimulated two-photon Raman
outcoupler for an atom laser using a full three-dimensional
description. A finite-temperature trapped Bose-condensed atomic
gas is treated self-consistently by the Hartree-Fock-Bogoliubov
equations. The model is closely related to a recent experiment on
optical outcoupling [E.W. Hagley {\it et al.}, Science {\bf 283},
1706 (1999)]. We analyze the momentum distribution of the output
atoms and show how the output beam may be used as a probe of the
quantum state for the trapped atomic gas and how it could be
engineered and controlled in a nonlinear way.
\end{abstract} 

The generation of Bose-Einstein condensates (BECs) of weakly interacting atomic gases has the potential to open a whole new area in atom optics; coherent atom optics. To do this, a well-controlled source of coherent matter waves needs to be developed. Rudimentary coherent matter wave sources (``atom lasers") have been engineered experimentally \cite{MEW97,HAG99}, but there is a need for sustained development. In this Letter we present a quantum-mechanical, three-dimensional (3D), {\it ab initio} treatment of a pulsed Raman outcoupler for a BEC. We demonstrate how the output of the atom laser can be engineered and how it may be used as a probe of the quantum and thermal properties of the BEC. We use a well-tested self-consistent field-theoretical treatment to describe the finite-temperature, multi-mode BEC (the analogy of the cavity in a conventional laser), and well controlled and detailed approximations
for the output coupling to a quasicontinuum of states.

A fundamental difference between optical and atom lasers is that atoms interact while photons do not. This significantly complicates the theoretical analysis of atom lasers, as compared to optical lasers. Studies using classical mean fields (e.g.\ Gross-Pitaevskii equation) disregard thermal fluctuations, decoherence, and information about quantum statistics. Due to the inevitable computational difficulty of a complete 3D analysis, there has not existed before a rigorous 3D, multi-mode, finite-temperature description for an outcoupler of an atom laser, despite a notable research activity on atom laser output theory in recent years \cite{MOY97,MOY99,STE98,EDW99,BAN99,JAP99,JAC99}.

We study a pulsed outcoupling which is amenable to a simpler theoretical description than a general continuous coupling, since it allows us to reach the Markovian limit and to ignore any nontrivial coherences between the trapped and free modes. A non-Markovian output may emerge in the continuous case \cite{MOY99,JAC99}, which, in a multimode description for the trapped gas, can lead to very complicated nonequilibrium dynamics.

We start from the Hartree-Fock-Bogoliubov (HFB) \cite{Allan,HUT97}
field theory for the Bose-condensed
atoms, harmonically trapped in internal level $|c\>$, in
thermodynamic equilibrium. The Hamiltonian density for the trapped
atomic gas alone reads
\beq
\Hc_1=\psi_c^\dagger (H^{(c)}_{\rm
cm}-\mu^{(c)}) \psi_c+{g_c\over2} \psi_c^\dagger\psi_c^\dagger
\psi_c\psi_c\,,
\eeq
where $H^{(c)}_{\rm cm}\equiv -\hbar^2{\bbox
\nabla}^2/(2m)+m\omega^2r^2/2$ is the center-of-mass
Hamiltonian in an isotropic trap with frequency $\omega$.
The interaction strength $g_c\equiv 4\pi a_c \hbar^2/m$ is given
in terms of the $s$-wave scattering length $a_c$, and $\mu^{(c)}$
denotes the chemical potential. In the HFB theory we decompose the condensate and the noncondensate fractions into mutually orthogonal subspaces and expand the field operator
\beq
\psi_c(\rv)=\Phi(\rv)\alpha_b+\sum_{n}[u_n(\rv)\alpha_n-v_n^*(\rv)
\alpha_n^\dagger]\,,\label{field}
\eeq
where the quasiparticle creation and annihilation operators
$\alpha^\dagger_n$ and $\alpha_n$ obey the standard bosonic commutation
relations \cite{Allan}. The BEC wave function $\Phi(\rv)$ satisfies the
generalized Gross-Pitaevskii equation with the thermal population
acting as an external potential \cite{HUT97}. We self-consistently
solve for the mode functions $u_n$ and $v_n$ and the BEC density
profile (as explained in detail in Ref.~\cite{HUT97}) in such a
way that the Hamiltonian ${\cal H}_1$ is diagonal in the Popov
approximation (where two-particle correlations due to the anomalous
average of the fluctuating field operator are neglected).
The HFB-Popov treatment has been exceedingly successful in describing the properties of BECs in both isotropic traps \cite{HUT97} and, at relatively low temperatures, also in anisotropic traps \cite{HUT98}.

We study the outcoupling of the trapped quantum degenerate
Bose-Einstein gas from internal sublevel $|c\rangle=|g,m\rangle$
to an untrapped level $|o\>=|g,m''\rangle$. We concentrate on a
situation where the atoms in level $|o\>$ do not experience a
magnetic trapping potential. As in the NIST experiments
\cite{HAG99} we consider a two-photon output coupling via two
noncopropagating lasers: Level $|c\>$ is optically coupled to an
electronically excited state $|e\rangle=|e,m'\rangle$ by the laser
field $\Ev_{1}$ with the frequency $\Omega_1$, and state $|e\>$ is
coupled to level $|o\>$ by the driving field $\Ev_{2}$ with the
frequency $\Omega_2$. The atomic transition frequencies for
$|c\>\rightarrow |e\>$ and $|e\>\rightarrow |o\>$ are $\omega_{1}$
and $\omega_{2}$ ($\omega_{oc}\equiv \omega_{1}-\omega_{2}$),
respectively. The lasers are off-resonant, so that the population
in level $|e\>$ remains small and spontaneous emission may be
ignored. In the rotating wave approximation (rwa) it is useful to
define the slowly varying field operator $\tilde{\psi}_{o} =
e^{i(\Omega_1-\Omega_2) t}\psi_{o}$ for level $|o\>$ in the
Heisenberg picture and the positive frequency components
$\tilde{\bf E}^+_j = e^{i\Omega_j t}{\bf E}^+_j$, where $\tilde{\bf E}^+_j \equiv\pol_j{\cal E}_j(\rv t)/2$. In the general formalism we allow for the lasers to be focused at the center of the trap, although the main results in this paper are obtained for nonfocused lasers with a Gaussian time profile:
${\cal E}_j(\rv t) \propto e^{-t^2/(2\beta^2)} e^{i {\bf k}_{j}\cdot \rv }$. We set the positive $z$ axis to be the direction of the momentum kick ($\qv\equiv\kv_{1}-\kv_{2}$). For pulsed outcoupling with strong momentum kick perpendicular to gravity \cite{HAG99}, gravity is not expected to play as important role as in a rf output \cite{MEW97}. Nevertheless, in the diagnostics of the trapped BEC, it would be advantageous to cancel completely the effects of gravity. We assume this is done, e.g., by means of a pulsed magnetic field gradient or an optical potential during the output pulse and ignore gravity in the following analysis.

We adiabatically eliminate the excited state and keep only the
terms of first order in $1/\Delta_1$, the inverse of the
one-photon detuning $\Delta_1\equiv \omega_{1}-\Omega_1$. With
the outcoupling the Hamiltonian density is ${\cal
H}_1+{\cal H}_2$, where
\bea
\lefteqn{{\cal H}_2 = \tilde\psi^\dagger_{o}
(H^{(o)}_{\rm cm}-\mu^{(o)}-\hbar\delta_2+\hbar\delta_{oc})
\tilde\psi_{o}+g_{oc}\psi_c^\dagger
\tilde\psi^\dagger_{o}\tilde\psi_{o}\psi_c }\nonumber\\
&&\mbox{}+{g_{o}\over2}
\tilde\psi^\dagger_{o}\tilde\psi^\dagger_{o}
\tilde\psi_{o}\tilde\psi_{o} -\left(\hbar\kappa
\tilde\psi^\dagger_{o} \psi_{c}+{\rm H.c.}\right)
-\psi^\dagger_{c}\hbar\delta_1\psi_{c} \,, \label{eq:HDN3}
\eea
and $\delta_{oc}\equiv\omega_{oc}-\Omega_1+\Omega_2$ is the
two-photon detuning, $H^{(o)}_{\rm cm}=-\hbar^2{\bbox
\nabla}^2/(2m)$, and the light-induced level shifts $\delta_i$ and
the Rabi coupling $\kappa$ are defined by
\beq
\delta_i(\rv t)={|{\cal E}_i(\rv t)|^2d_{i}^2\over 2\hbar^2\Delta_1},\quad
\kappa(\rv t)= {{\cal E}_1 (\rv t){\cal E}^*_2 (\rv
t)d_{1}d_{2}\over 2\hbar^2\Delta_1}\,.
\label{eq:para}
\eeq
Here the reduced dipole matrix element $d_{1}$ ($d_2$) for the atomic
transition $|c\>\rightarrow |e\>$ ($|o\>\rightarrow |e\>$) also
contains the corresponding nonvanishing Clebsch-Gordan
coefficient.

By inserting Eq.~{(\ref{field})} into the coupling terms
proportional to $\kappa$ in Eq.~{(\ref{eq:HDN3})} we immediately
observe that, despite introducing the rwa, the creation of one
output atom may be associated with the annihilation
($\psi_o^\dagger \alpha_n$) {\it or} creation ($\psi_o^\dagger
\alpha_n^\dagger$) of a quasiparticle in a trap \cite{JAP99}.
Consequently, even at $T=0$, and in the weak coupling limit, the
outcoupling of an atom can create elementary excitations in the
quasiparticle vacuum demonstrating how the full multimode approach
with quasiparticle excitations is unavoidable in obtaining the
complete description of the outcoupling process.

We consider
a low density for the outcoupled atoms
so that $\mu^{(o)}\simeq0$ and we may ignore the
term proportional to $g_o$ in Eq.~(\ref{eq:HDN3}). Moreover, we
expand the output matter field in the plane wave basis
$\psi_o(\rv)=\sum_\kv \<\rv|\kv\> o_\kv$, with $\<\rv|\kv\> \equiv
V^{-1/2}\exp{(i\kv\cdot\rv)}$, where $o_\kv$ is the annihilation
operator for the mode $\kv$. We may then obtain the Heisenberg
equation of motion for slowly varying output mode operators
$\tilde{o}_\kv\equiv\exp{[i(\epsilon_\kv+\omega_{oc})t]} o_\kv$
with $\epsilon_\kv=\hbar \kv^2/(2m)$:
\beq
\dot{\tilde{o}}_\kv =
i\sum_{\kv'} \xi_{\kv
\kv'}(t)\tilde{o}_{\kv'}+i\nu_{\psi_c}^{(\kv)}(t)
e^{i(\epsilon_\kv+\delta_{oc})t}\,. \label{eqnm}
\eeq
We have defined the coupling matrix elements, analogous to the
Franck-Condon factors, by $\nu_{f}^{(\kv)}(t)\equiv
\<\kv|\kappa(\rv t)| f\>$ and $ \xi_{\kv \kv'}(t)\equiv
\<\kv|[\delta_2(\rv t)-g_{oc}n^{(c)}(\rv)/\hbar]|\kv'\>
e^{i(\epsilon_\kv-\epsilon_{\kv'})t} $. The collisional
interaction between the trapped and output atoms is treated
semiclassically: $\psi_o^\dagger\psi_o\psi_c^\dagger\psi_c \rightarrow \psi_o^\dagger\psi_o n^{(c)}$.

For the Markov and Born approximation to be valid we ignore the depletion of the trapped modes, requiring that the light only weakly perturbs them, $\delta_1\lesssim \omega$, and that we only couple out a small fraction of atoms, $\beta\kappa\ll 1$. In the first Born approximation we write $\alpha_n(t)\simeq \alpha_n(-\infty) e^{i\omega_nt}$. For a continuous outcoupling the Markovian limit would be more restrictive, since then the atoms should also leave the interaction region (the trapped gas of size $R$) much faster than any timescale $\tau$ required to form coherences between the trapped and free modes, indicating $\tau\hbar q/m\gg R$. The timescale $\tau$ determined by the collision rate $\tau\sim \hbar/(g_{oc}n^{(c)})$ or by the coupling rate $1/\kappa$ necessitates a very strong laser kick $q$.

In Eq.~{(\ref{eqnm})} $\xi_{\kv\kv'}$
describes the coupling between the different output modes as a
result of the laser focusing and the interactions with the trapped
atoms. For short pulses, $\beta\ll \hbar/(g_{oc}n^{(c)})$, (or when $\hbar^2 q/m\gg R g_{oc}n^{(c)}$) the atomic collisions between the trapped and free atoms can be ignored during the laser coupling, and we may set $g_{oc}\simeq 0$ in $\xi_{\kv\kv'}$ \cite{com}. Moreover, if also the spatial variation of the laser intensity is negligible over the atomic cloud (nonfocused case), the coupling between the different $\kv$ modes vanishes due to the orthogonality of the plane waves and we obtain
$\xi_{\kv\kv'}\simeq \delta_2 \delta(\kv-\kv')$. In this case the
solution of Eq.~{(\ref{eqnm})} reads:
\bea
\lefteqn{\lim_{t\rightarrow\infty} \tilde{o}_\kv
(t)/(2i\pi)=\sum_n \left[ \alpha_n \tilde\nu_{u_n}^{(\kv)}
\delta_{2\over
\beta}(\epsilon_\kv+\delta_{oc}'-\omega_n)\right.}\nonumber\\
&&\mbox{}\left.-\alpha^\dagger_n \tilde\nu_{v^*_n}^{(\kv)}
\delta_{2\over\beta}(\epsilon_\kv+\delta_{oc}'+\omega_n)\right]
+\alpha_b \tilde\nu_{\Phi}^{(\kv)}
\delta_{2\over\beta}(\epsilon_\kv+\delta_{oc}')\,.\label{pop}
\eea
Here the operators $\alpha_n$ and $\alpha^\dagger_n$ are evaluated
at the thermodynamic equilibrium $t\rightarrow -\infty$ and
$\delta_{oc}'\equiv \delta_{oc}-\mu^{(c)}/\hbar$. The
dynamics of the coupling matrix elements has been factored out: $\tilde\nu_f^{(\kv)}\equiv
e^{t^2/\beta^2}\nu_f^{(\kv)}(t)=\nu_f^{(\kv)}(0)$. We also
introduced a delta-function with a finite width $\Gamma$:
$
\delta_\Gamma (x)\equiv \pi^{-1/2} \Gamma^{-1} \exp{(-x^2/\Gamma^2)}
$.
In the limit of long pulses $\beta\rightarrow\infty$ the width is zero and $\delta_\Gamma(x)$ becomes the Dirac delta function.

In Eq.~{(\ref{pop})} we have ignored the light-induced level
shifts in the delta functions by assuming a weak coupling
$\delta_i \ll \omega_n,\epsilon_\kv$. The physical interpretation
of Eq.~{(\ref{pop})} is straightforward. The expression represents
the output mode amplitudes after the laser pulses, $\beta\lesssim
t$. The conservation of the momentum is expressed
in the coupling matrix elements $\nu$ between the modes. The
integrand of $\nu$ is nonnegligible over some finite volume, with
the characteristic length scale $R_0$, resulting in a momentum
uncertainty $\Delta k\sim 1/R_0$. The delta functions dictate the
energy conservation. Due to the finite duration of the pulse the
delta functions have acquired a finite width, inversely
proportional to the pulse length; again a direct consequence of
the Heisenberg uncertainty relation. The sign of the quasiparticle
energy in the delta functions is different in the terms
proportional to $u_n$ and $v_n^*$ corresponding to the
annihilation and creation of an excitation associated with the
outcoupling of an atom. In the limit of a short pulse,
$\beta^2(\epsilon_\kv+\delta_{oc}'\pm\omega_n)^2/4 \ll 1$, the
delta functions become constant, and we obtain $ \tilde{o}_\kv
(\infty)\simeq i\beta \pi^{1/2}\nu^{(\kv)}_{\psi_c}(0)$, where
$\psi_c$ is evaluated in the initial state $t\rightarrow -\infty$.
Then the output after the pulse is proportional to a one-particle correlation function of the trapped atoms:
\beq
\<o_\kv^\dagger o_\kv\>\propto\int d^3r d^3r'\, e^{i(\kv-\qv)\cdot (\rv-\rv')} \< \psi_c^\dagger(\rv)\psi_c(\rv')\>\,.
\label{dens}
\eeq
For a homogeneous BEC the output mode density can measure a Fourier component of the standard first-order spatial coherence function. Equation (\ref{dens}) should be contrasted to the Bragg spectroscopy which probes {\it two-particle} correlations of the trapped atoms \cite{STA99}.

In a more general case the output mode density is obtained from Eq.~{(\ref{pop})}. If the collisions between the trapped and the
output atoms cannot be ignored, or if the laser is focused, the
result (\ref{pop}) is no longer valid, and we need to solve the
coupled set of differential equations (\ref{eqnm}). In general,
evaluating the output field boils down to calculating the dynamics and the coupling matrix elements $\nu^{(\kv)}_{u_n}$,
etc. In a spherically symmetric trap we may expand the mode
functions $u_n(\rv)=\sum_{\bar{n}\bar{l}\bar{m}} C_{\bar{n}\bar{l}\bar{m}}
\phi_{\bar{n}\bar{l}}(r)Y_{\bar{l}\bar{m}}(\theta,\varphi)$ [similarly for $v_n(\rv)$] in terms of the radially
symmetric wave functions $\phi_{\bar{n}\bar{l}}(r)$ and the
spherical harmonics $Y_{\bar{l}\bar{m}}(\theta,\varphi)$. The
symmetry is simplified in the case of nonfocused lasers or when
the focused lasers are parallel. Then the outcoupling is
cylindrically symmetric around the direction of the momentum kick
($z$ axis). Consequently, the Rabi frequencies $\kappa(r,\theta)$
are independent of the polar angle $\varphi$ which can be
integrated analytically: $ \kappa_{\bar{n}\bar{l}\bar{m}}\equiv
\<\kv|\kappa (r,\theta)|u_n\> =
\<k_\rho,k_z|\kappa(r,\theta)|\phi_{\bar{n}\bar{l}}\>$, where we may write
$\<\rho,z|k_\rho,k_z\> \equiv
{\cal A}_{\bar{l}\bar{m}} P^{|\bar{m}|}_{\bar{l}}(z/r) J_{\bar{m}}(k_\rho
\rho)e^{ik_z z}$, with $\rho^2=x^2+y^2$. Here $J_k$ denotes the $k$th order Bessel function of the first kind, $P^m_l$ the associated Legendre function, and ${\cal A}_{\bar{l}\bar{m}}$ complex coefficients. As
a result, the momentum distribution of the atom laser is
characterized by the two components, $k_z$ and $k_\rho$,
representing the momentum along the laser kick and
perpendicular to it.

In the numerical calculations we consider an outcoupling process with nonfocused lasers from an isotropic 3D trap with the trapping frequency $\omega=2\pi\times 100$Hz by evaluating the operator expectation values for the output mode density from Eq.~{(\ref{pop})}. The BEC profile, the thermal population, the excitation frequencies $\omega_n$, and the quasiparticle mode functions, $u_n$ and $v_n$, are solved for 10000 $^{23}$Na atoms with the scattering length $a=2.8$nm and the coupling strength $\bar{\kappa}\equiv\kappa(0,0)=0.05\omega$. This involves a self-consistent evaluation of about 1000 excitation frequencies and the respective mode functions \cite{HUT97}. The $(2\times2)$D coupling matrix elements $\kappa_{\bar{n}\bar{l}\bar{m}}$ are numerically integrated for ($\bar{n},\bar{l},\bar{m}$). With 10000 atoms the number of outcoupled atoms in some examples remains small, but it becomes experimentally reasonable if we scale the results for experiments on several millions of atoms.

\begin{figure}
\vbox{ \vspace{-3mm}\hbox{ \epsfig{width=2.8truecm,file=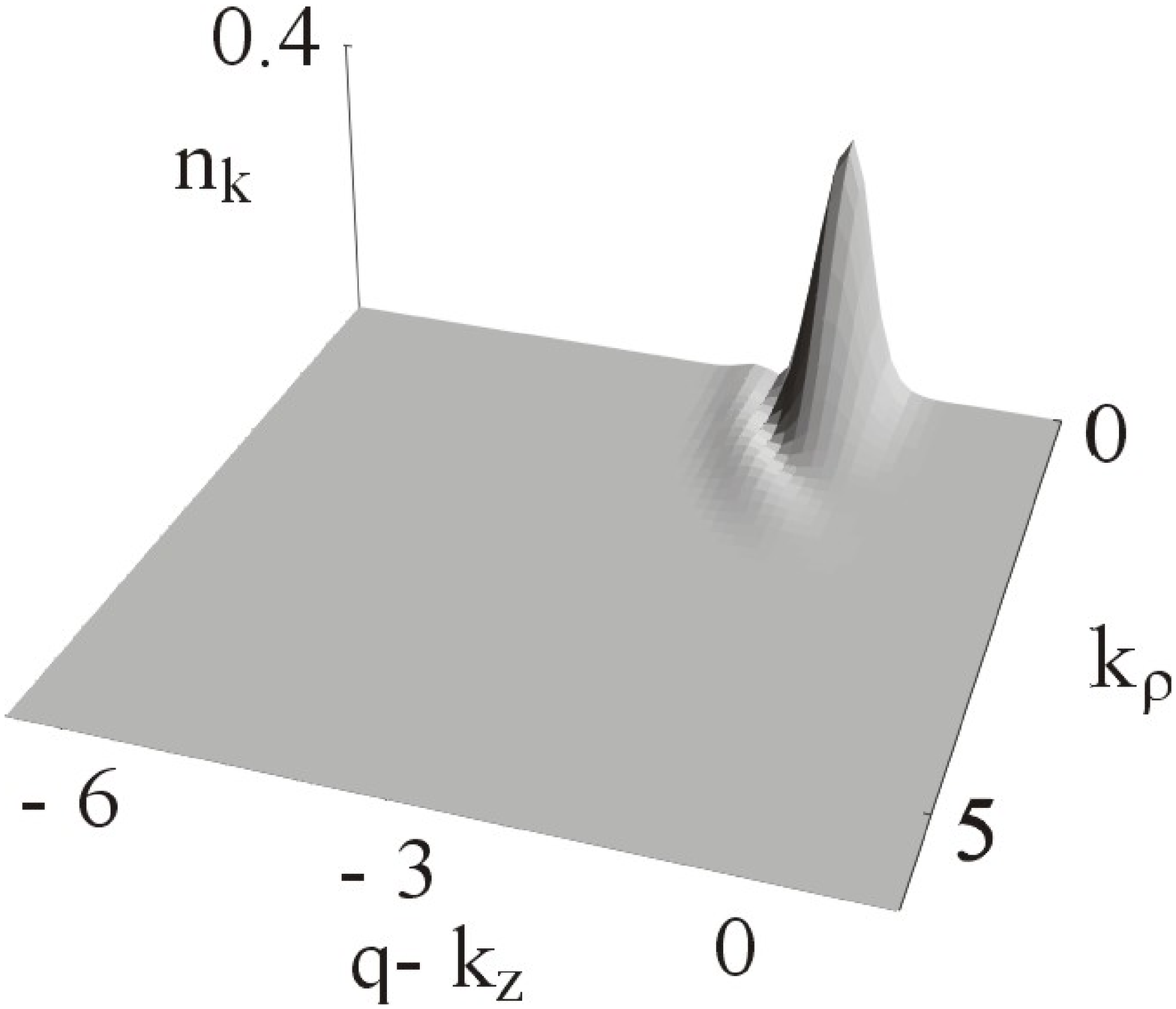}
\epsfig{width=2.8truecm,file=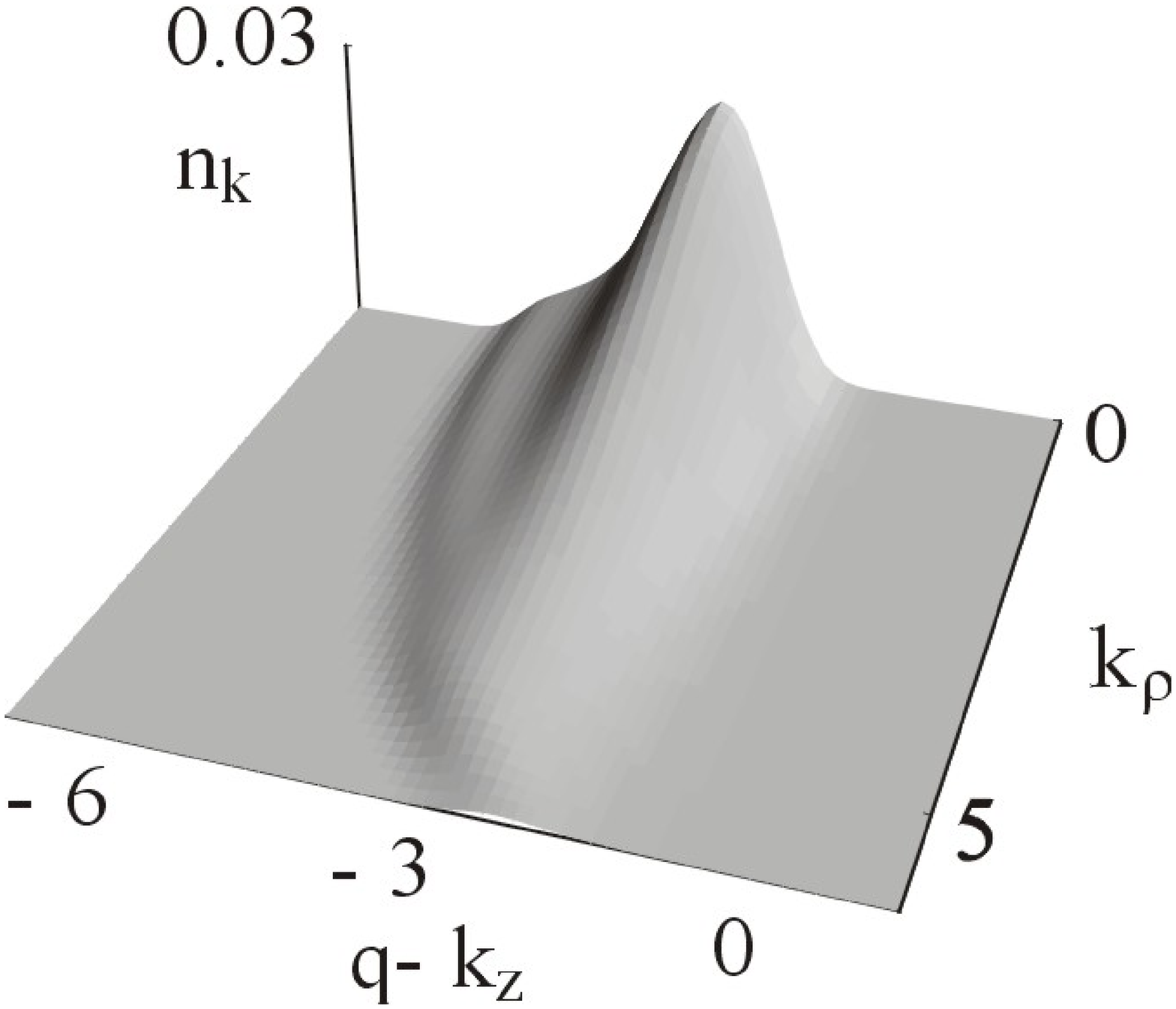}
\epsfig{width=2.8truecm,file=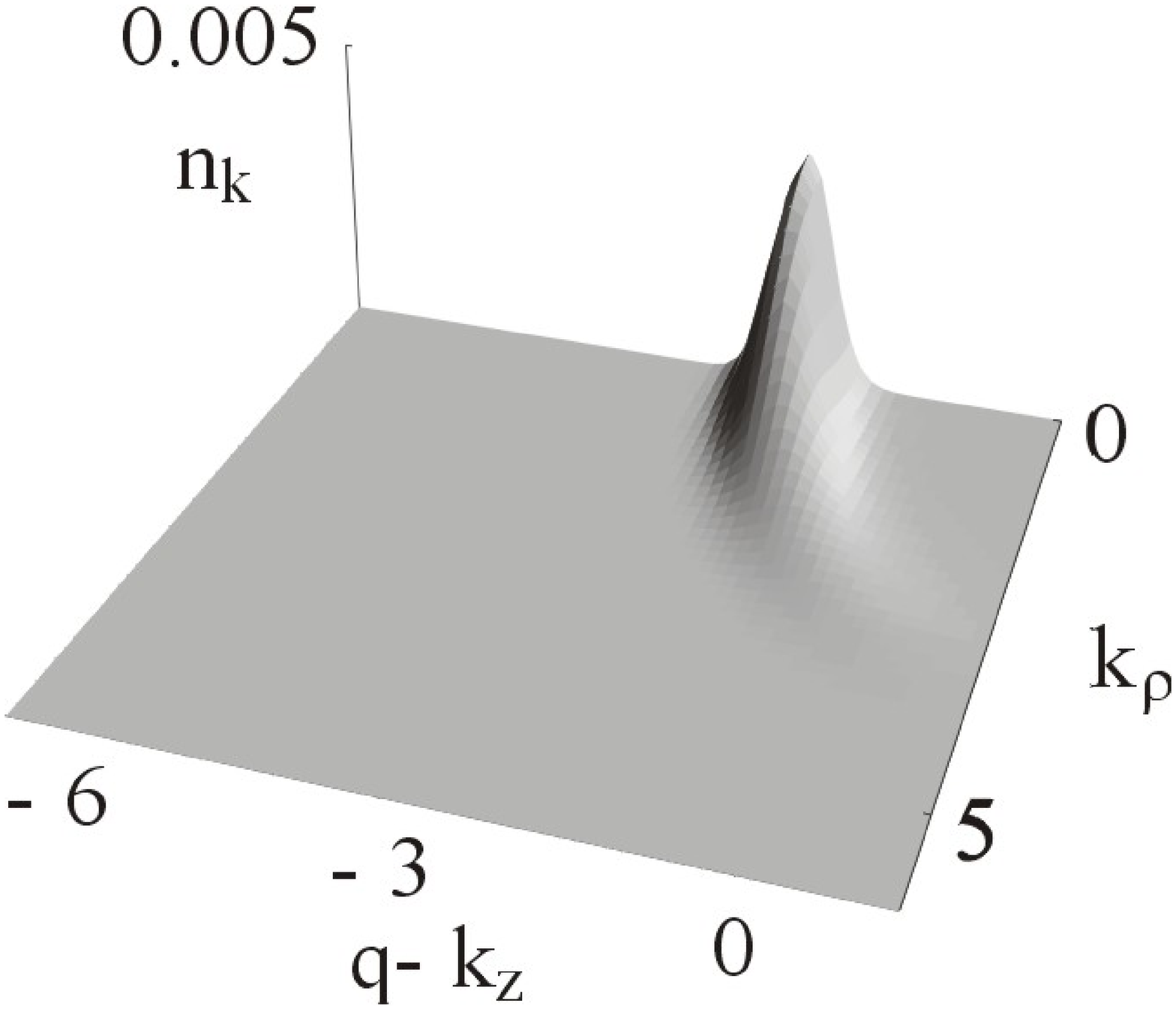}
}}\caption{The momentum distribution of the outcoupled atoms along the direction of the momentum kick ($k_z$) and perpendicular to it ($k_\rho$) representing the coherent output from the BEC (left), the annihilation (middle), and the creation (right) of a quasiparticle in the trap. The corresponding processes are peaked at $k_z l\simeq 5.6$, $6.8$, and $5.9$ resulting in a spatial separation of the propagating output beams.
 } \label{fig1}
\end{figure}

At $T=60$nK we obtain 7070 BEC atoms in the trap with the
corresponding 2930 atoms in the noncondensate fraction. In
Fig.~\ref{fig1} we display the momentum distribution of the
outcoupled atoms corresponding to the different coupling processes
proportional to $\Phi$, $u_n$, and $v_n^*$. We have chosen the
detuning $\delta_{oc}'=-20\omega$, the pulse length
$\beta=0.5/\omega$, and the momentum kick $q=0.1\times
4\pi/\lambda$, where $\lambda=589$nm represents the wavelength of
light for typical optical transitions in $^{23}$Na. The momentum
distributions corresponding to the three different mechanisms are
clearly separated due to the different resonance energies. This
results in the spatial separation of the three output fields,
which could be used to probe the quantum properties of the trapped
gas.

Energy conservation, determined by Eq.~{(\ref{pop})}, is optimized
when $\epsilon_q+\epsilon_{k_z'}+\epsilon_{k_\rho}-\hbar q k_z'/m
+\delta_{oc}'=\pm \omega_n$, where $k_z'$ denotes the momentum of
the atoms along the laser kick relative to the momentum kick of
the lasers, $k_z'\equiv q-k_z$. With $q=0$ the output is
spherically symmetric. Energy conservation is optimized for
different values of $k_z'$ for the output processes proportional
to $\Phi$, $u_n$, or $v_n^*$, with $-\omega_n$ and $\omega_n$
representing $v_n^*$ and $u_n$, respectively. The energy condition
is in general matched at two different values of $k_z'$. In the
case of Fig.~\ref{fig1} we obtain for the BEC $k_z'\simeq -1.9/l$
and $k_z'\simeq 11/l$ [$l\equiv (\hbar/m\omega)^{1/2}$],
representing the atoms flying in the opposite directions from the
trap. However, in general, the two energy maxima can represent
very different occupation numbers due to the different strengths
of the coupling matrix elements. In particular, momentum
conservation for the coherent coupling from the BEC is optimized
when $k_z'=k_\rho=0$, i.e., when $\epsilon_q =-\delta_{oc}'$. In
Fig.~\ref{fig1} only the first maximum is displayed which is
approximately three orders of magnitude larger than the second
one. Nevertheless, it is fascinating that, with large BECs, we
still find parameter regimes where the output from a BEC mode is
represented by two distinguishable fragments with an observable
number of atoms flying in the direction opposite to the laser
kick, indicating a coherent splitting of the BEC output.

\begin{figure}
\vbox{ \vspace{-3mm}\hbox{ \epsfig{width=2.8truecm,file=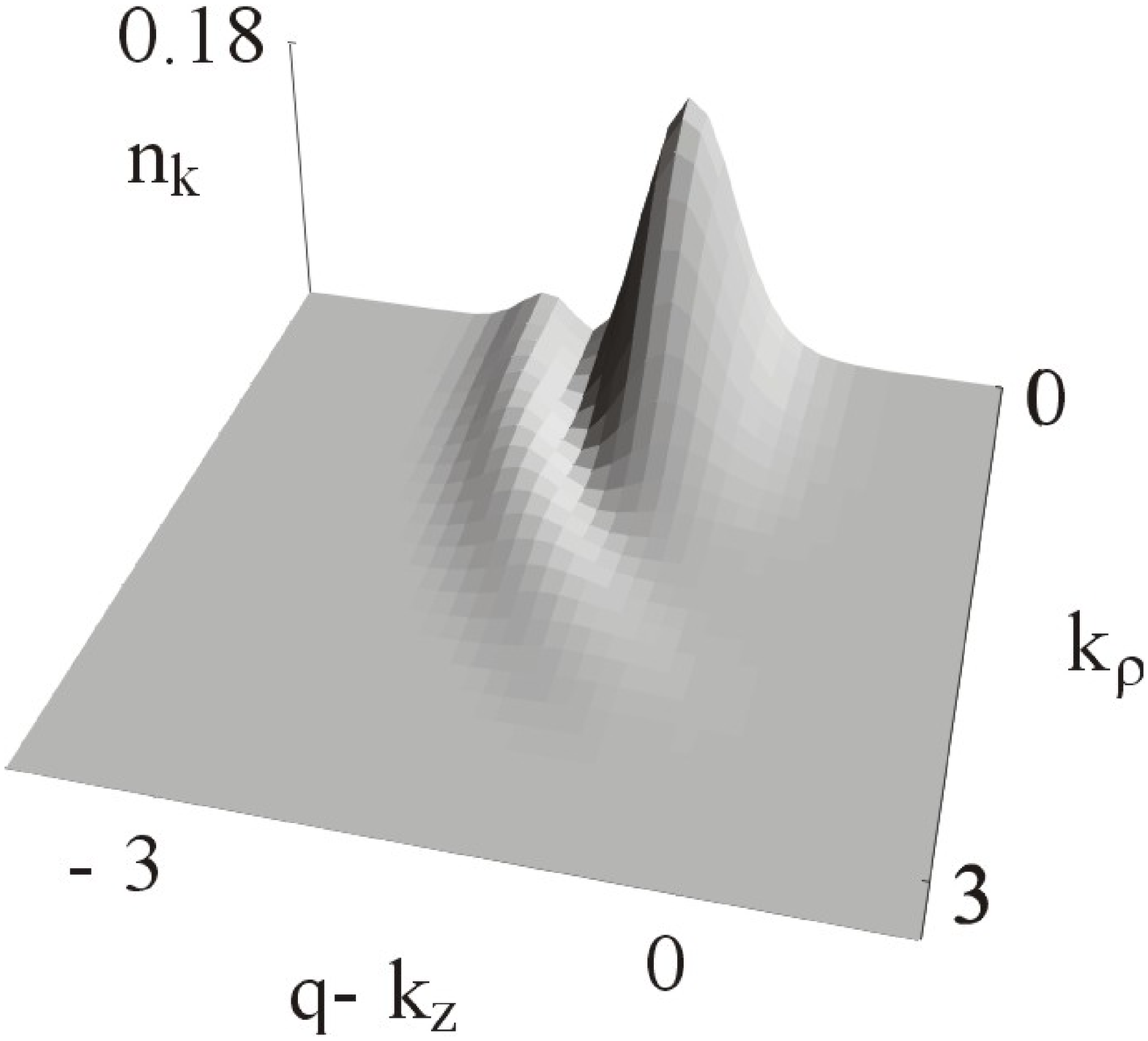}
\epsfig{width=2.8truecm,file=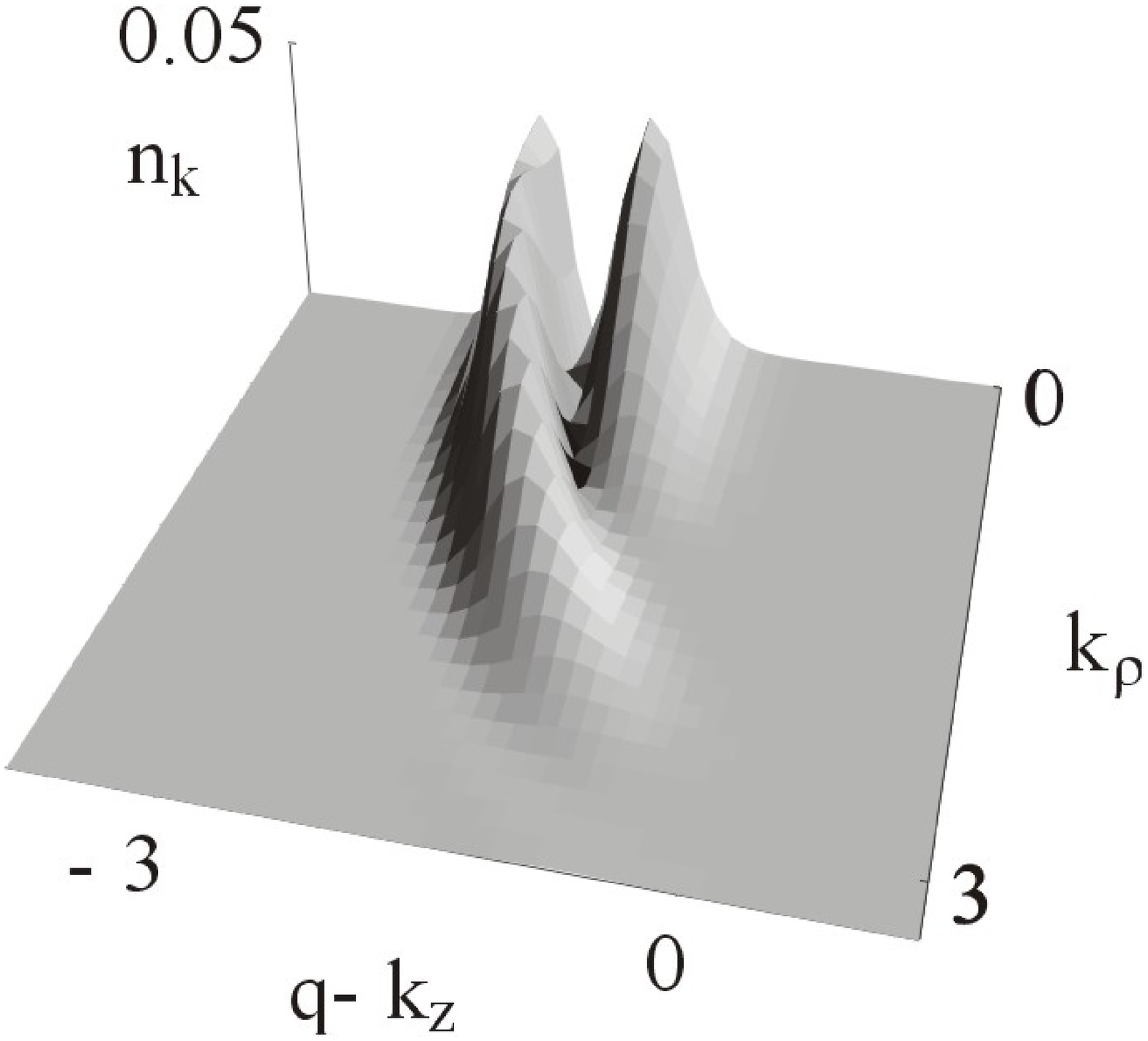}
\epsfig{width=2.8truecm,file=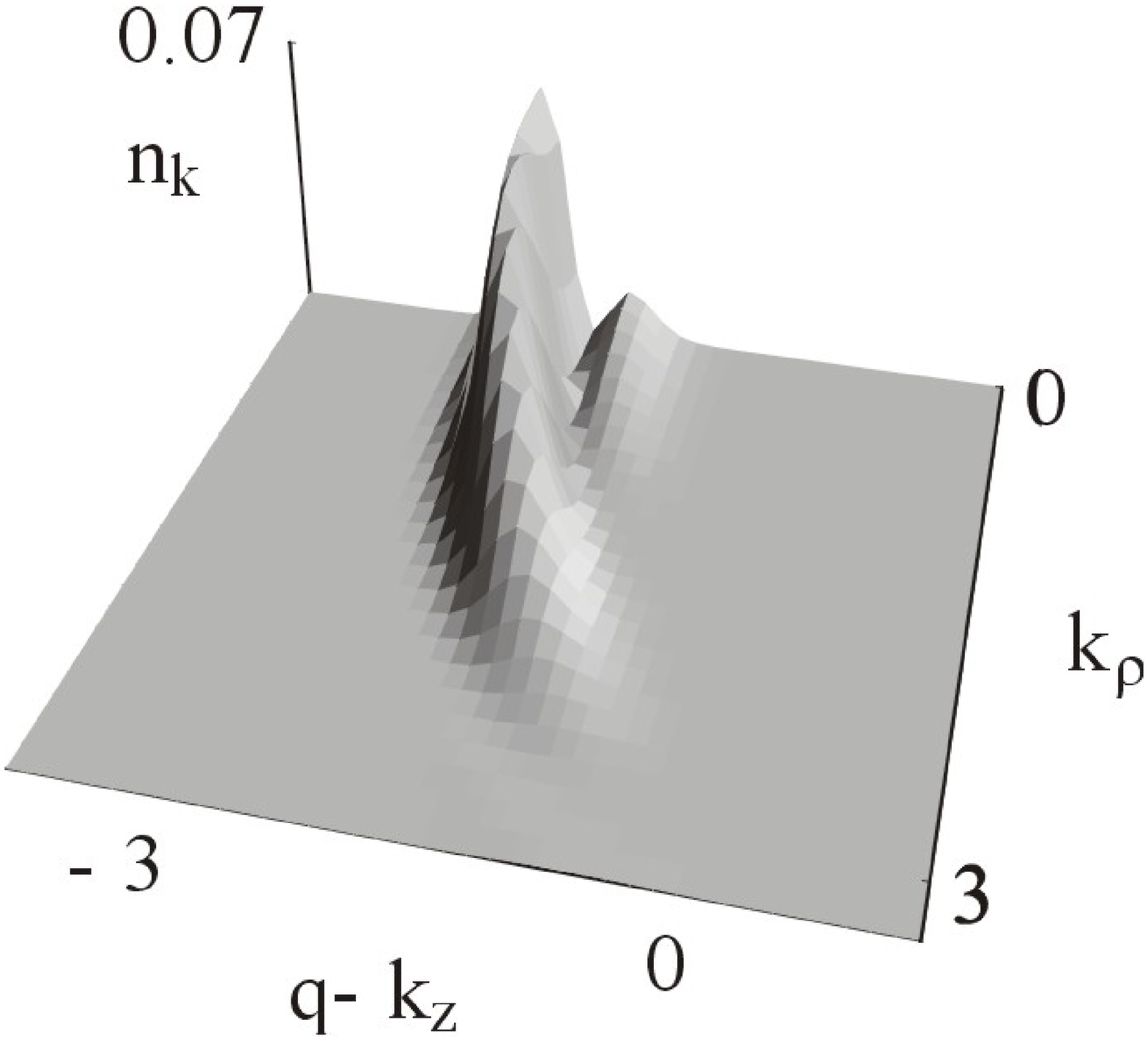}
}}\caption{
The momentum distribution of the coherently outcoupled atoms from the BEC for three different values of the pulse duration $\beta$, with fixed $\delta_{oc}'=-10\omega$, $N=7073$, and $q=0.05\times 4\pi/\lambda$. With $\beta\omega=0.6$ (left) momentum conservation is well-optimized representing a peak at $k_z\simeq q\simeq 2.2/l$. For longer pulses energy conservation becomes better optimized resulting in arising second peak at $k_z\simeq 2q$. With $\beta\omega=0.8$ (middle) the two peaks exhibit equal hights. With $\beta\omega=1.0$ (right) energy conservation dominates.
 } \label{fig2}
\end{figure}

Whether the output distribution is more determined by the
optimized energy or momentum conservation depends on the spatial
confinement of the trap and on the duration of the laser pulses
$\beta$. In Fig.~\ref{fig2} we show the momentum distribution for
atoms coupled out of the BEC mode for three different values of
$\beta$ describing a transition from well-optimized momentum
conservation to well-optimized energy conservation.

\begin{figure}
\vbox{ \vspace{-3mm}\hbox{ \epsfig{width=4.3truecm,file=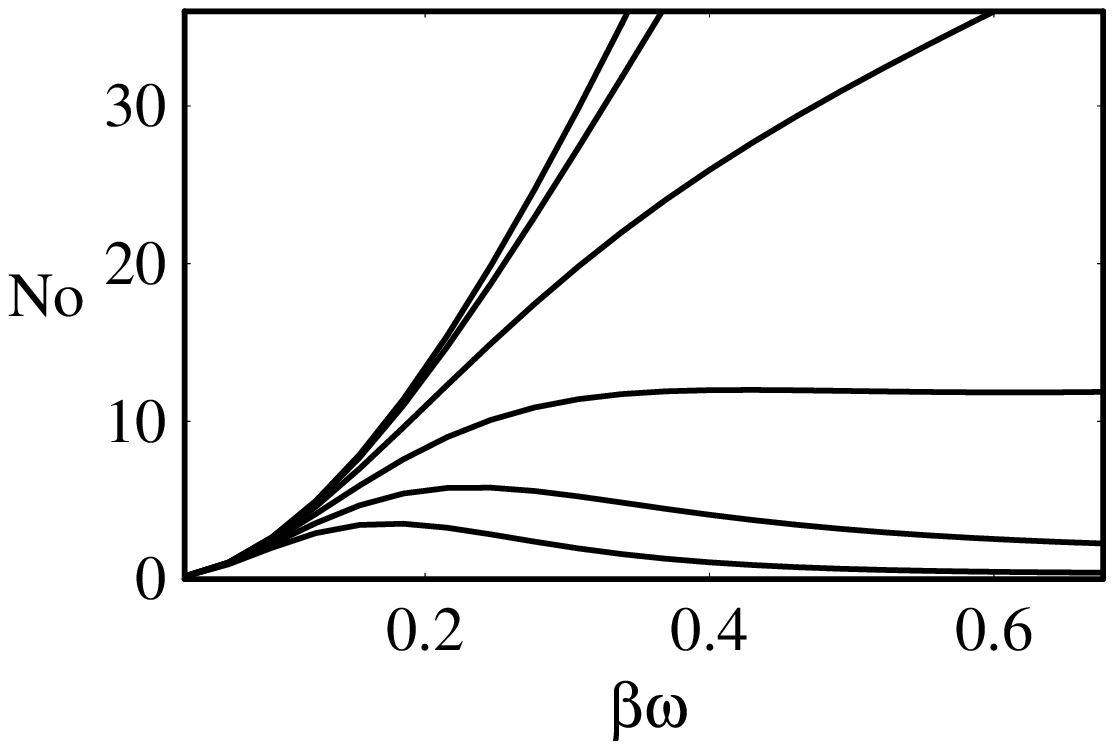}
 \epsfig{width=4.3truecm,file=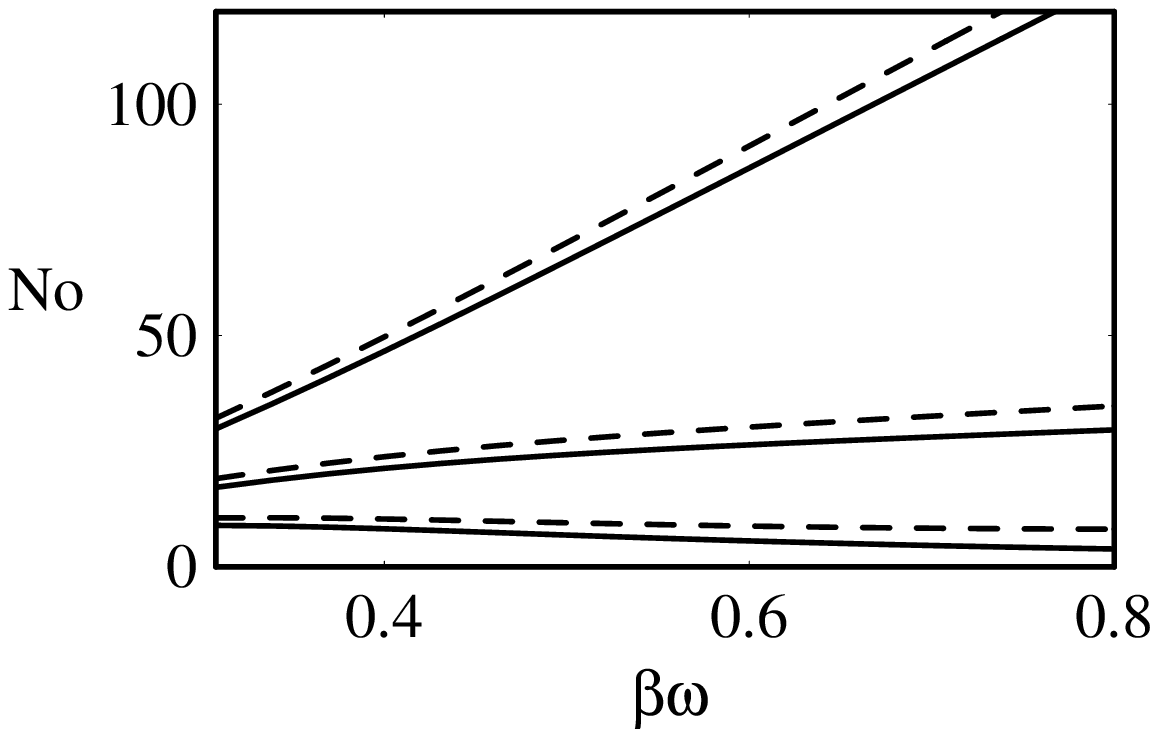}
}}\caption{
The number of coherently outcoupled atoms from the BEC mode (left) as a function of the pulse length $\beta$. Beginning from the lowest curve, the value of the detuning changes evenly from $\delta_{oc}'/\omega=-20$ to $-10$ ($\epsilon_q\simeq 10\omega$). The coherent coupling dominates the output close to $T=0$. The right plot shows the coherently outcoupled atoms (solid lines) and the total number of outcoupled atoms at $T=60$nK (dashed lines) for $\delta_{oc}'/\omega=-16$, $-14$, and $-10$.
} \label{fig3}
\end{figure}

As the final example we investigate the number of outcoupled atoms
$N_o$. In Fig.~\ref{fig3} we show the number of coherently and
incoherently outcoupled atoms as a function of the pulse length
for $q=0.1\times 4\pi/\lambda$ and for different values for the
detuning. We observe that $N_o$ may also {\it decrease} with {\it
increasing} $\beta$. This seemingly counter-intuitive result can
be explained by noting that the uncertainty of energy conservation
decreases with increasing pulse length. The finite-width delta
function in Eq.~{(\ref{pop})} then approaches the Dirac delta
function and a nonresonant coupling becomes suppressed. Even very
small changes of the laser detuning $\delta_{oc}$ can result in
large modifications of the output. Therefore, e.g., externally
induced level shifts could possibly provide a strongly nonlinear
control of the atom laser output.

The essential features of the coherent output in Fig.~\ref{fig3} may be understood by deriving an approximate expression for the total number of atoms, coherently coupled out from the BEC, by assuming a Gaussian density profile for the BEC with the width $R$ along the laser kick:
\beq
N_o\sim N \bar{\kappa}^2 \beta^2 \chi
\exp{\[ -\chi^2 \beta^2 (\epsilon_q+\delta_{oc})^2/2 \]}\,,
\label{outrate}
\eeq
where $\chi^2\equiv (q R)^2/[(q R)^2+2(\beta\epsilon_q)^2]$. In the derivation of Eq.~{(\ref{outrate})} we assumed that the width of the BEC density profile perpendicular to the laser kick satisfies $R_\rho/l \gg \beta\omega \chi$. The approximate expression (\ref{outrate}) describes well the functional dependence of the output. In Fig.~\ref{fig3} the best fit is obtained by choosing $R\sim 0.5 R_{TF}$ for $\delta_{oc}'<-10\omega$ and $R\sim 0.6 R_{TF}$ for $\delta_{oc}'>-10\omega$, where the Thomas-Fermi radius $R_{TF}\equiv l(15N a_c/l)^{1/5}$. For short pulses $2 (\beta \epsilon_q)^2\ll (qR)^2$, we obtain $\chi\simeq 1$, and the output depends quadratically on the pulse length $N_0\sim N \bar{\kappa}^2\beta^2$. For long pulses, $2 (\beta \epsilon_q)^2\gg (qR)^2$, the output is linear with $\beta$: $N_o\sim N \bar{\kappa}^2\beta q R/(\sqrt{2}\epsilon_q)\exp{\{-[qR(1+\delta_{oc}/\epsilon_q)/2]^2\} }$.
There is also an interesting intermediate regime, where $(qR)^2\sim 2(\beta\epsilon_q)^2$ and $\chi^2\sim1/2$. Provided that the coupling is sufficiently off-resonant, so that we simultaneously have $2\lesssim \beta |\epsilon_q +\delta_{oc}|$, the exponential factor in Eq.~{(\ref{outrate})} becomes important, and the number of output atoms may rapidly decrease as a function of the pulse length for fixed coupling strength. This behavior results from energy conservation and the very narrow energy width of the BEC and is a consequence of the genuine quantum nature of the BEC, as compared to thermal atomic ensembles with a broad energy distribution.

This research was supported by EPSRC, A. von Humboldt Foundation,
the European Commission, and the Marsden Fund of the Royal Society of New Zealand.

\begin{references}
\bibitem{com} Alternatively, the interactions may also be considered negligible compared to the laser kick if $\hbar\epsilon_q\gg g_{oc}n^{(c)}$.
\end{references}
\end{document}